\documentclass[10pt,a4paper,english,journal]{IEEEtran}
\usepackage[latin9]{inputenc}
\usepackage{graphicx}

\makeatletter

\pdfpageheight\paperheight
\pdfpagewidth\paperwidth

\DeclareTextSymbolDefault{\textquotedbl}{T1}
\providecommand{\tabularnewline}{\\}

\usepackage[left=1.62cm,right=1.62cm,top=1.9cm, bottom=4.3cm]{geometry}
\usepackage{lipsum}
\usepackage{microtype}
\usepackage{xcolor,soul}
\sethlcolor{lightgray}

\usepackage[justification=centering]{caption}

\makeatother

\usepackage{babel}
\begin{document}
\title{Evolution of 3GPP Standards Towards True Extended Reality (XR) Support
in 6G Networks}
\author{\IEEEauthorblockN{Ali A. Esswie, and Morris Repeta\\
Standards \& Industry Forums, Advanced Wireless Technology, Dell Technologies }}

\maketitle
$\pagenumbering{gobble}$
\begin{abstract}
Extended reality (XR) is a key innovation of 5G-advanced and beyond
networks. The diverse XR use-cases, including virtual reality, augmented
reality, and mixed reality, transform the way humans interact with
surrounding environments. Thus, XR technology enables true immersive
experiences of novel services spanning, e.g., e-commerce, healthcare,
and education, respectively. However, the efficient support of XR
services over existing and future cellular systems is highly challenging
and requires multiple radio design improvements, due to the unique
XR traffic and performance characteristics. Thus, this article surveys
the state-of-art 3GPP standardization activities (release-18) for
integrating the XR service class into the 5G-advanced specifications,
highlighting the major XR performance challenges. Furthermore, the
paper introduces valuable insights and research directions for supporting
true XR services over the next-generation 6G networks, where multiple
novel radio design mindsets and protocol enhancements are proposed
and evaluated using extensive system level simulations, including
solutions for application-native dynamic performance reporting, traffic-dependent
control channel design, collaborative device aggregation for XR capacity
boosting and offload, respectively. 

\end{abstract}

\section{Introduction}

Extended reality (XR) is one of the most promising service classes
of future cellular systems. XR use-cases, including virtual reality
(VR), augmented reality (AR), mixed reality, complemented by advanced
cloud gaming (CG), offer immersive human experiences for collaborative
gaming, telepresence and hybrid workplaces {[}1{]}. However, the success
of the mass XR adoption depends on the efficient support of future
cellular systems to the challenging XR quality of service (QoS) and
computing demands. Therefore, XR requires substantial specification
efforts from various standardization bodies, and specially, the 3rd
generation partnership project (3GPP), due to its unique performance
and traffic characteristics {[}2{]}. 

Particularly, the XR QoS requirements and traffic characteristics
do not fit within the existing 5G-native QoS classes such as the ultra-reliable
and low-latency communications (URLLC) and enhanced mobile broadband
(eMBB). That is, XR applications demand an eMBB-alike extreme capacity
with URLLC-alike tight radio latency and reliability targets {[}3{]}.
This makes fulfilling the XR radio performance targets highly challenging
and many of the existing radio operations (e.g., radio resource scheduling
and radio performance reporting) become sub-optimal due to the native
design of the 5G radio, which is optimized towards URLLC-eMBB coexistence
{[}4{]}. 

Accordingly, 3GPP partners have been extensively studying XR service
and traffic characteristics, and therefore, identifying respective
radio challenges and potential solutions for XR capacity and latency
optimization. Specifically, the standardization of XR via 5G has started
in 2016, where the 3GPP service and system aspects (SA1) working group
(WG) initiated the specification of the 5G target service requirements
for high-capacity and low-latency applications {[}5{]}. Following
SA conclusions, the radio access network (RAN1) WG has completed a
study item, during 5G release-17, on the traffic characteristics of
the various XR applications {[}6{]}. Later, during the current 5G
release-18 (to be tentatively completed by 2023), the corresponding
normative work has started within both RAN1 and RAN2 WGs {[}7{]},
in order to specify the needed radio enhancements to support basic
XR services, including aspects for capacity enhancements, scheduling
enhancements, and XR-specific power saving {[}8-10{]}, respectively. 

In this paper, we survey the state-of-the-art 3GPP specification efforts
for XR service support via the 5G new radio. The major XR-specific
performance challenges over 5G are reported, and which are currently
under 3GPP study and improvement. This includes multiple radio aspects
as XR radio resource management, XR traffic Jitter adaptation, and
the XR multi-flow multi-priority traffic, respectively. In the second
part of the paper, we propose and evaluate multiple novel radio designs
and solutions for a more efficient XR support over 5G-adavnced and
beyond radio interfaces, and which transform the current 3GPP specification
mindset. Those are, an application-dependent radio performance management,
traffic-aware control channel design, and collaborative multi-device
aggregation for XR capacity offload and boosting, respectively. Finally,
the paper offers insightful conclusions on how the 3GPP standardization
is changing its basic specification mindset to efficiently support
XR services over future cellular systems. 

The paper is organized as follows. Section II presents the current
XR performance challenges, which are under ongoing 3GPP study during
the active release-18. Section III discusses the proposed solutions
and corresponding simulation results for true XR service support over
future cellular networks. Conclusions and recommendations are drawn
in Section IV.

\section{XR Performance Challenges in 5G New Radio}

XR applications demand a highly perceived data rate with strict packet
delay budget (PDB) and reliability levels, respectively, placing it
between the eMBB and URLLC service classes. Moreover, XR has unique
traffic characteristics, which are dependent on the specific XR use-case
(where more than 20 XR use cases are identified {[}11{]} by 3GPP SA
WGs). The video streaming is the basic traffic type of XR applications
in the downlink direction (and in uplink direction as well for AR
applications). This leads to an enormous amount of XR traffic, available
for immediate scheduling, and which is associated with stringent radio
latency and reliability targets. Accordingly, the network spectral
efficiency is highly degraded due to the fundamental trade-off between
data rate, latency, and reliability. Subsequently, The vast XR traffic
characteristics and performance requirements make existing radio operations,
e.g., radio resource management, become sub-optimal to efficiently
integrate XR services as follows.

\subsection{Sub-Optimal XR Scheduling Procedures}

Scheduling is a vital radio function where the network dynamically
allocates the available radio resources to different active devices,
in order to meet their different QoS metrics. There are generally
two scheduling categories: Dynamic and semi-static scheduling, respectively
{[}12{]}, as follows. 
\begin{itemize}
\item \textbf{Dynamic scheduling}: it implies the network dynamically schedules
radio resources to devices when either downlink or uplink traffic
is available for transmission. This requires multiple exchanges of
control channel information, for the network to be aware of how much
uplink traffic is available at the devices, and/or for indicating
devices with the scheduled downlink/uplink resources and associated
transmission configurations, e.g., adopted modulation and coding schemes,
and the payload re-transmission modes.
\item \textbf{Semi-static scheduling}: namely as downlink semi-persistent
scheduling or uplink configured grant scheduling, it denotes that
network pre-schedules a certain amount of downlink or uplink resources
for different devices, and which are periodically repeated in the
time domain over a configured scheduling validity period. Accordingly,
the straightforward advantage of semi-static scheduling is avoiding
the transmission of multiple control channel information, which leaves
more valuable radio resources for useful payload transmissions. 
\end{itemize}
Dynamic scheduling is best utilized when the traffic inter-packet
arrival rate and traffic volume are not consistent in time, i.e.,
time-varying or sporadic; though, semi-static scheduling is appropriate
for almost periodic traffic arrivals with a pre-determined payload
size. The XR traffic characteristics have been deemed to fall in a
middle class between these two traffic classes. For instance, for
virtual reality services with view-port dependent streaming, traffic
volume and inter-packet arrivals are highly time-variant due to the
dynamic selection and switching of the adopted video codecs, even
for the same downlink or uplink traffic flow. However, for augmented
reality applications, downlink and uplink traffic are deterministic.
For instance, after each received downlink payload, an uplink control
update follows within a strict latency budget. 

Thus, with dynamic scheduling, XR traffic is independently scheduled
in downlink and uplink directions, with multiple control channel information
exchanges prior to the actual payload scheduling. This adapts the
resource scheduling to the frequent changes in the traffic arrival
rate, periodicity and volume; however, it leads to a significantly
degraded channel capacity and an increased packet buffering delay,
which may violate the stringent XR latency QoS. With semi-static scheduling,
periodic resource grants are allocated to XR devices to avoid the
repeated control message exchanges. However, due to the dynamic downlink
Jitter, which is an XR-application and XR-server specific, packet
availability in the time domain may deviate from the time instant
the semi-statistically granted resources become active, and thus,
clearly degrading the network spectral efficiency {[}13{]}. 

\subsection{Application-Unaware Radio Performance Reporting}

Since the era of the third generation (3G), the 3GPP mindset has dictated
specifying radio capabilities that meet the performance requirements
of target use cases and services, instead of specifying use-case-specific
solutions and procedures. Such mindset is making the efficient support
of XR services quite challenging. For instance, during an active session,
a device needs to continuously compile several radio measurement and
performance reports, and accordingly, send those back to the serving
network node. Examples are (a) the channel quality indication (CQI)
reporting {[}14{]} for indicating the network with the actual channel
and interference conditions, (b) buffer status reporting (BSR) {[}15{]},
where it indicates the network about the size of the buffered uplink
traffic, available at the device, and accordingly, the network can
schedule the sufficient amount of uplink resources, and (c) hybrid
automatic repeat request (HARQ) feedback reporting {[}16{]}, where
it signals the network whether the received downlink payload is successfully
decoded or not. 

Clearly, those performance reports are vital for the network to efficiently
schedule resources and to select the appropriate transmission configurations
for each traffic flow. However, particularly for XR services, and
due to application-unawareness of the performance reporting procedures,
those schemes may mislead the network and further influence a sub-optimal
resource scheduling as follows.
\begin{itemize}
\item \textbf{\textit{Application-outage-unaware HARQ feedback}}: HARQ reports
indicate the network whether formerly transmitted packets are successfully
decoded or not. Upon receiving a negative HARQ feedback, the network
triggers a packet re-transmission accordingly. For reducing the radio
trip latency, HARQ packet re-transmissions are always prioritized
for transmission over new packet arrivals. For a variety of XR applications,
not all traffic flows are equally important nor equally impacting
end user experience, e.g., view-port dependent XR streaming {[}6{]}.
Hence, the network scheduler may be dictated by non-important HARQ
packet re-transmissions, which are always prioritized over possibly
more vital new payload arrivals. 
\item \textbf{\textit{Application-independent BSR reporting precision}}:
BSR reporting is key for devices to signal the network nodes of the
size of their buffered uplink traffic. Accordingly, network can sufficiently
schedule the proper amount of uplink resources. BSR reporting follows
a predefined structure, where a BSR table is specified such that devices
select a row index from the BSR table which corresponds to a range
buffered traffic size, satisfying their actual buffered traffic volume.
The BSR quantization precision is predefined and designed in such
way to offer a higher quantization precision (with less quantization
errors) over the lower range of the traffic volume and a lower precision
over the higher range. The lower quantization precision denotes that
a single BSR index covers a large range of the buffered traffic volume.
Hence, the network shall not identify exactly how much traffic is
buffered at devices, leading to either over-scheduling or under-scheduling
of resources (and thus, a degraded spectral efficiency or increased
traffic buffering latency). However, such design fits perfectly within
existing URLLC-eMBB coexistence deployments. Specifically, the URLLC
traffic is always small sized, and therefore, utilizing the higher
BSR quantization precision over the lower traffic volume range, while
eMBB applications have no stringent latency targets, and thus, they
can efficiently coexist with under-scheduling network strategies,
i.e., requiring multiple scheduling grants for the entire buffer to
be transmitted over a longer duration. For XR services, the traffic
volume, inter-packet arrivals, frame size, and radio latency targets
(to be satisfied) are highly time variant and dependent on the XR
application itself. This leads the current BSR reporting procedure
to likely offer a higher BSR reporting precision over a buffered traffic
range that may not be experienced by the XR application, leading to
a poor uplink scheduling performance.
\end{itemize}

\subsection{Semi-Static Control Channel Design}

Control channels are critical to achieve reliable 5G communications
{[}17{]}. Exchange of control information, in both downlink and uplink
directions, serve multiple purposes including exchange of resource
allocation information, sharing of transmission configurations, and
carrying performance related key performance indicators (KPIs), respectively.
Thus, without a timely decoding of the control information, multiple
radio procedures are significantly disturbed, e.g., resource scheduling
and dynamic link adaptation. Hence, due to the vital impact of control
channels on the achievable radio latency and reliability, they are
designed in a semi-static and  extreme reliable way such that its
corresponding decoding probability is always maximized. That is, control
channels are statically modulated using a highly conservative modulation
scheme and with a strong channel coding rate, e.g., QPSK modulation
to increase the decoding ability, since without a successful decoding
of control channels, devices or the network may not even be aware
of, for instance, granted data resources or critical performance-related
KPIs. 

Due to the conservative modulation order and strong encoding, the
capacity of control channels is highly limited. Specifically, transmitting
a small amount of control information towards a single device takes
over a large set of valuable resources. With multiple simultaneously
active devices, where each must be assigned a dedicated control channel
for continuous monitoring, the control channel overhead can significantly
consume the network capacity, i.e., leaving less resources for useful
data transmissions. Accordingly, optimization of control channel capacity
is vital for enhancing the aggregate cell spectral efficiency. With
XR services, where an enormous amount of traffic volume per device
is available for transmission, resources consumed by control channels
must be highly limited such that to leave the remaining resources
for the data transmissions; however, due to the semi-static design
of the current 3GPP control channel, control channel capacity can
not be flexibly adapted without impacting either the total network
spectral efficiency (when increasing the control channel resources)
or the XR critical QoS targets (when reducing the amount of control
channel resources, and therefore, devices may exhibit delayed or even
canceled control channel transmissions, leading to a degraded latency
performance). Therefore, the current 3GPP control channel design mindset
needs to be upgraded to introduce capacity- and reliability-adaptive
control channels, which are dependent on the diverse importance of
the XR traffic flows, for which the control information is transmitted.

Therefore, those radio inefficiencies lead to a significantly degraded
network capacity. During the current 3GPP release-18 study on XR {[}3{]},
the user satisfaction ratio per cell is adopted as a key XR KPI, and
is defined as the number of XR users, compared to the total number
of active XR devices, for which their latency and capacity QoS targets
are satisfied. As depicted by the exemplary system-level evaluation
in Fig. 1 {[}18{]}, the XR user satisfaction ratio is presented for
three XR use-cases, namely AR, VR, and CG, with 30 and 45 Mbps as
the target data rate. The deployment setup is: Dense Urban of 21 macro
cells, each equipped with 64 Tx antennas, 100 MHz bandwidth per cell.
Obviously, the 95\%ile of the satisfied XR user ratio is highly limited
for the three XR services, despite the large 100 MHz bandwidth allocation
per cell. Specifically, the 95\%ile network capacity of VR/AR with
a target rate of 30 Mbps is 4.2 users per cell while it drops to 1.6
users per cell with 45 Mbps. Also, increasing the number of active
XR devices clearly reduces the effective satisfied user ratio due
to the additional buffering delay. 

\begin{figure}
\begin{centering}
\includegraphics[scale=0.46]{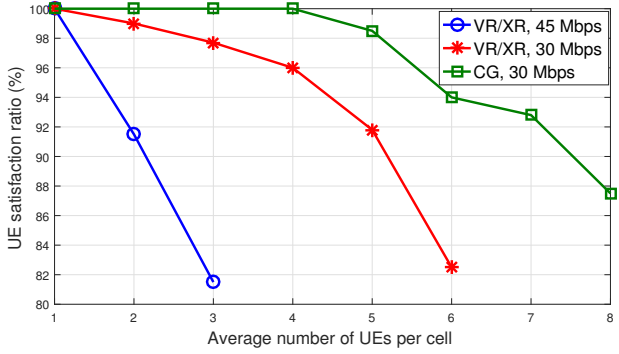}
\par\end{centering}
\centering{}\caption{User satisfaction ratio of AR, VR and CG.}
\end{figure}

\begin{figure}
\begin{centering}
\includegraphics[scale=0.42]{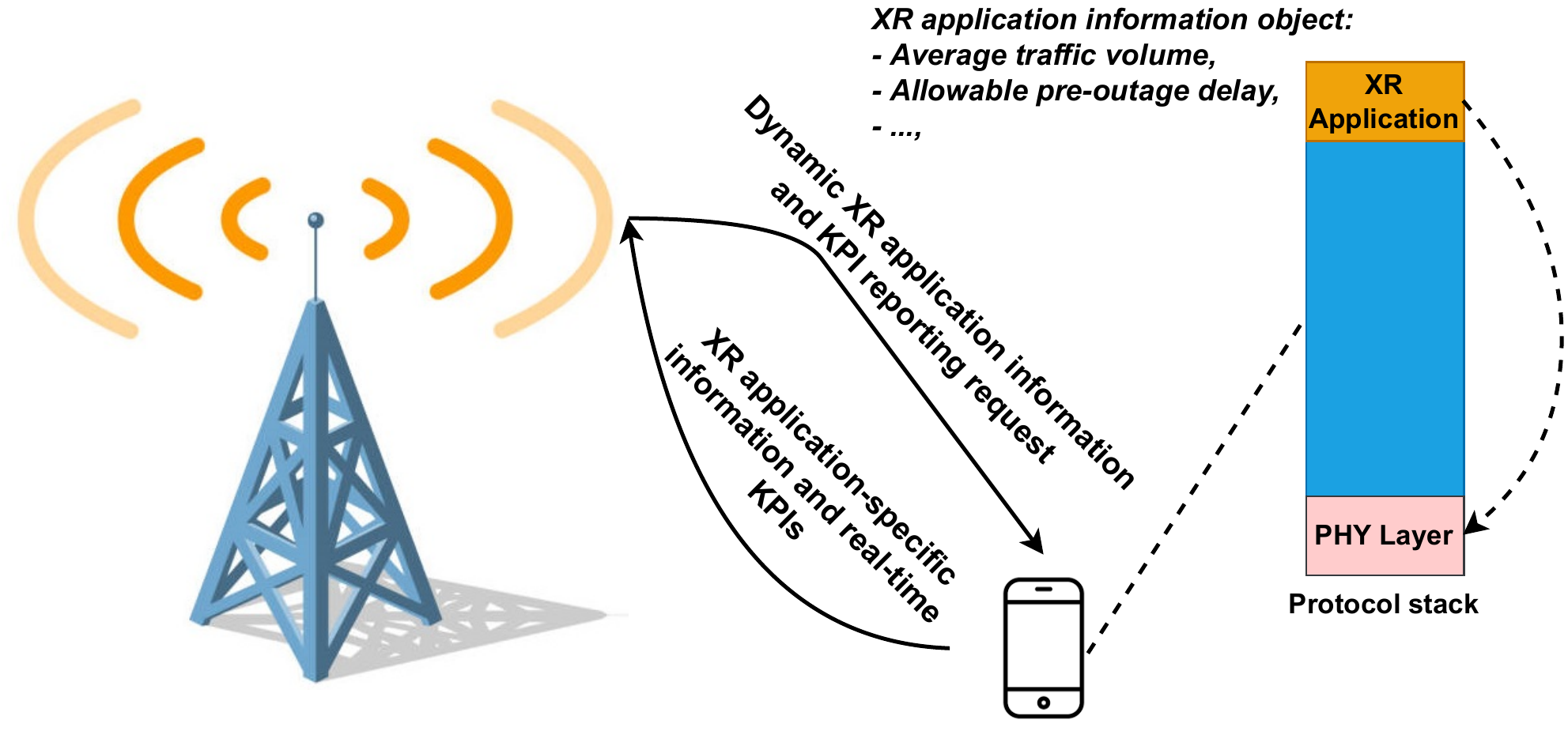}
\par\end{centering}
\centering{}\caption{Proposed application-driven performance reporting.}
\end{figure}

\section{Evolution of 3GPP Standards Towards 6G XR: Insights and Research
Directions}

We address the XR-specific performance challenges, detailed in Section
II, and propose three novel radio design improvements, towards 5G-advanced
and beyond systems, to enable an efficient support of the stringent
XR services. 

\subsection{Application-Aware Performance Reporting }

Application-awareness, augmented in cellular radio performance reporting,
is not aligned with current 3GPP specification mindset, where 3GPP
bodies typically seek to specify performance KPIs and reporting procedures
that are related to basic radio capabilities or services, rather than
an application. However, based on the performance findings of the
3GPP release-17 study on XR, such mindset is deemed to critically
limit the flexibility and efficiency of cellular systems to support
the emerging XR services, where various XR applications adopt different
traffic profiles, QoS demands, and device behaviors, respectively.
With existing application-agnostic performance reporting, the network
fails to perform proper resource scheduling and adopt adequate transmission
configurations for the active XR application.

Therefore, as depicted by Fig. 2, we propose introducing a radio application
programmable interface (R-API) for carrying XR application-specific
information and real-time KPIs. Therefore, the network becomes able
to dynamically request a certain active device to calculate and report
a novel application KPI. Examples can be diverse and not limited to
XR services, e.g., average traffic volume, and real-time outage latency
among each successive uplink control/update transmission and downlink
reception. Therefore, the network can fine tune its scheduling policies
and transmission configurations specifically to the real-time needs
of the stringent applications. Aside from existing standardized KPI
reporting, the introduced R-API procedure allows the network and devices
to dynamically coordinate on calculating and reporting new and non-standardized
performance and application KPIs, \textit{on-the-go}, depending on
the processing ability of both entities. 

As an exemplary implementation of the proposed R-API, an application-driven
BSR precision reporting is depicted in Fig. 3. Unlike the current
fixed-precision BSR reporting procedures (Detailed in Section II),
proposed BSR precision adaptation allows the network to position a
more precise BSR reporting over the real-time buffered traffic volume
that is expected from the end application, and hence, performing more
resource- and latency-friendly uplink scheduling. Specifically, based
on the R-API concept, the network dynamically requests a device to
track, calculate and report the real-time average and standard deviation
of the XR application uplink traffic volume, and hence, the BSR tables
can adopt more precise BSR quantization steps, only over the indicated
traffic volume range, i.e., saving the unnecessary BSR reporting overhead
from the high precision BSR reporting over the non-probable traffic
range, while always satisfying high precision reporting over the expected
application traffic volume. This way, an application-specific BSR
reporting is utilized, instead of existing network-specific BSR standardized
reporting. 
\begin{figure}
\begin{centering}
\includegraphics[scale=0.42]{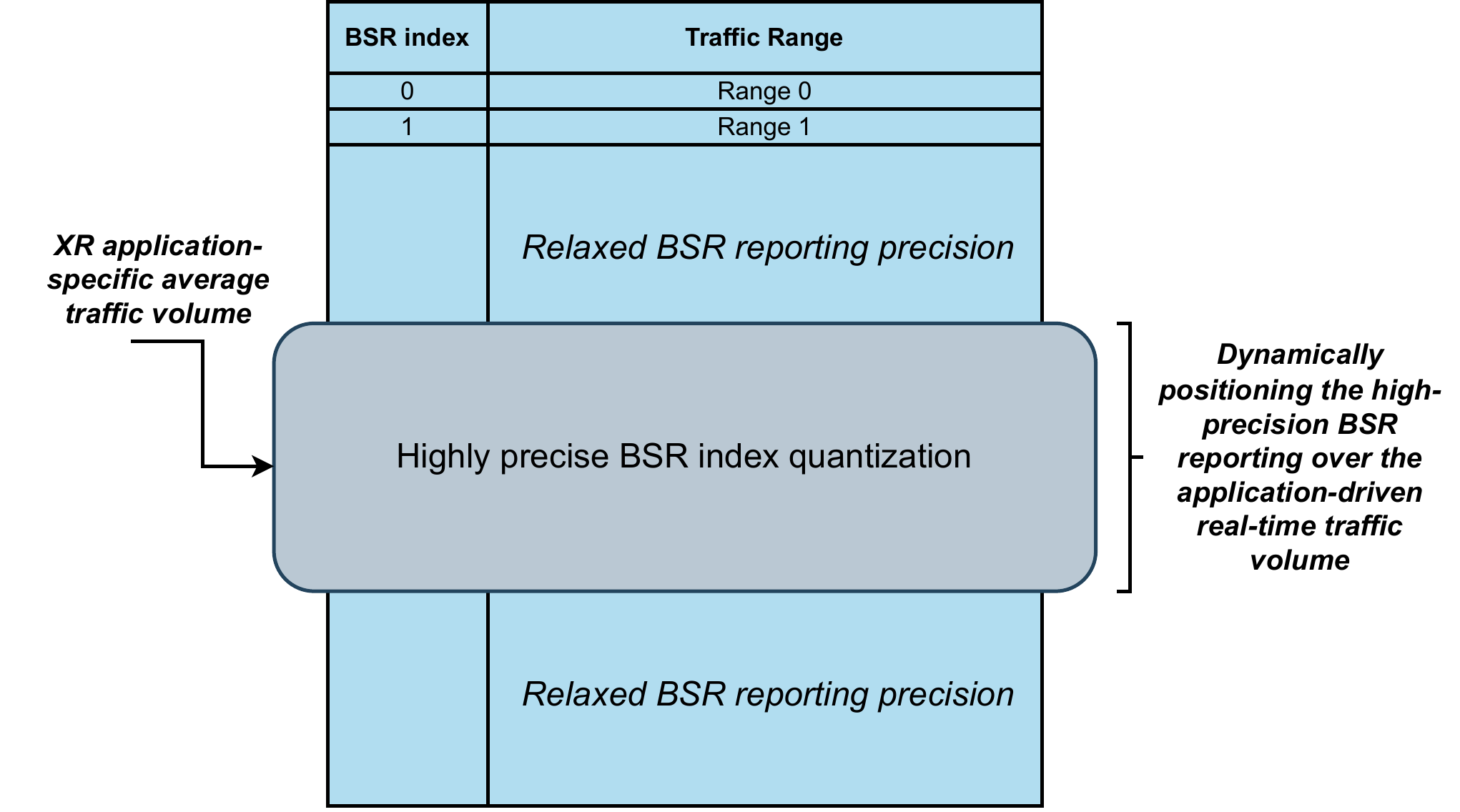}
\par\end{centering}
\centering{}\caption{Dynamic precision BSR design.}
\end{figure}

To validate the performance of the proposed application-driven BSR
design, we develop extensive system level simulations, following the
major 3GPP simulation guidelines. Table I lists the major simulation
settings. A 3GPP-compliant 5G Macro deployment is simulated, with
21 cells, each is equipped with 8 transmit antennas. There are 10
UEs on average per cell, each with 2 receive antennas. The FTP3 traffic
model is adopted in both downlink and uplink directions, respectively,
with sporadic payload arrival rate (i.e., following a Poisson distribution)
and equal offered traffic load, following the latest 3GPP RAN1 methodology
for XR evaluation {[}6{]}. The effective time-variant packet arrival
rate emulates both high-rate video and low-rate audio XR streams.
Herein, device mobility is only considered for channel fading calculations;
however, no inter-cell handover is utilized, i.e., handover cut-off
delays are ruled out in this evaluation. The uplink performance indicators
including the BSR and CQI indications are calculated and assumed available
to the cells for downlink and uplink scheduling. 

\begin{table}
\caption{{\small{}Simulation parameters.}}

\centering{}%
\begin{tabular}{c|c}
\hline 
Parameter & Value\tabularnewline
\hline 
Deployment & 3GPP Macro, 7 sites, 21 cells, Wraparound\tabularnewline
\hline 
Inter-site distance & 500 meters\tabularnewline
\hline 
Channel model & 3GPP-3D channel\tabularnewline
\hline 
Spectrum & 2.4 GHz\tabularnewline
\hline 
Bandwidth & 40 MHz\tabularnewline
\hline 
Antenna setup & Tx = 8, Rx = 2\tabularnewline
\hline 
Traffic model & FTP3, packet size = 1 Mbytes\tabularnewline
\hline 
UE dropping & Uniform, 10 UEs/cell\tabularnewline
\hline 
Link adaptation & Adaptive MCS\tabularnewline
\hline 
TTI & 7-OFDM symbols\tabularnewline
\hline 
Numerology & 30 KHz\tabularnewline
\hline 
UE speed & 3 Km/hr\tabularnewline
\hline 
DL and UL scheduler & Proportional fair \tabularnewline
\hline 
CQI mode & Sub-band of 8 PRBs\tabularnewline
\hline 
Duplexing mode & FDD\tabularnewline
\hline 
DL and UL receiver & Linear MMSE\tabularnewline
\hline 
Transmit power & BS: 43 dBm, UE: 23 dBm\tabularnewline
\hline 
HARQ mode & Async with Chase Combining\tabularnewline
\hline 
\end{tabular}
\end{table}

Fig. 4 depicts the empirical cumulative distribution function (ECDF)
of the achievable uplink throughput per UE, with three BSR reporting
schemes: (1) the standardized 3GPP 8-bit BSR reporting, (2) the proposed
application-driven BSR design, with 8-bits of reporting overhead,
and (3) a reference BSR design, with 10-bits of reporting overhead,
respectively. The three BSR schemes cover the same buffered traffic
range, i.e., from 0 Bytes to 81.3 Mbytes; although, with various quantization
steps and precision. Specifically, the 8-bit 3GPP reporting scheme
follows a predefined BSR quantization, where the BSR precision is
higher over the lower traffic range and looser over the larger traffic
range, e.g., for a buffered traffic volume below 100 bytes, a single
BSR index indicates a range of traffic that is only couple of bytes
wide; however, for buffered traffic volume above 15 Mbytes, a single
BSR index indicates a range of traffic that is a couple of Mbytes
wide. The 10-bit reference BSR design implies a fixed and more precise
BSR quantization over the entire supported traffic volume as the 3GPP
8-bit BSR scheme, i.e., $BSR_{step}=(81.3*1000)/(2^{10})=79.3$ Kbytes
over the entire BSR table. 

Finally, the proposed BSR scheme adopts variable BSR quantization
steps over the different traffic volume ranges, where the steps are
determined based on the proposed R-API and the additional XR application-specific
assistance signaling of the expected average traffic volume. For the
sake of evaluation, we assume a generated application-specific average
traffic volume of 10 Mbytes per second (with a packet inter-arrival
rate of 10 packets per second), and accordingly, the BSR quantization
precision/step is refined over the traffic volume range of $10\pm\alpha$,
where $\alpha=5$ Mbytes is a design parameter and determines the
dynamic traffic range of the refined BSR steps. Thus, the BSR quantization
precision is 3x refined over the determined range than the entire
BSR table, i.e., the BSR step size over the application-specific traffic
range (5 to 15 Mbytes) is 3x less than the step sizes adopted over
the remaining traffic range in the BSR table. 

As can be observed from Fig. 4, the proposed BSR design outperforms
the current 3GPP BSR reporting, while exhibiting the same BSR reporting
overhead of 8-bits. That is, over the 90\%ile level, a 66.6\% increase
in the uplink throughput per UE is achieved with the proposed BSR
reporting design, compared to the 3GPP BSR design. This is attributed
to the proposed BSR framework refining the BSR quantization precision
only over the dynamic traffic range of the XR application, and thus,
the accuracy of the uplink scheduling is improved by avoiding the
sub-optimal over-scheduling, leaving more resources for other devices.
The reference 10-bit BSR design offers the best achievable throughput
per device due to the always-refined BSR precision regardless of the
application traffic characteristics; however, at the expense of the
increased reporting overhead which limits the number of supported
devices.

\begin{figure}
\begin{centering}
\includegraphics[scale=0.51]{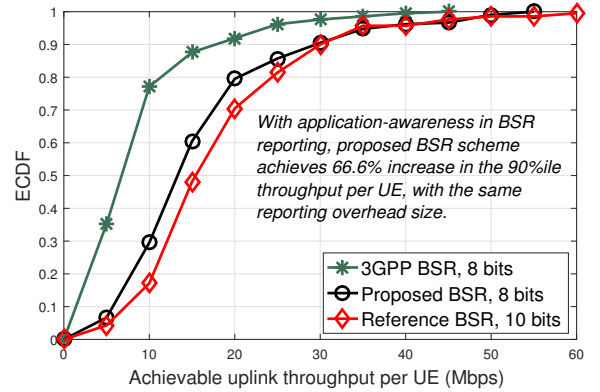}
\par\end{centering}
\centering{}\caption{ECDF of the uplink throughput per UE (Mbps).}
\end{figure}

\subsection{Traffic-Aware Control Channel Design}

The capacity of control channels is always deemed to be a critical
limitation of the stringent cellular services. This is due to several
fundamental factors as: (a) the semi-static and conservative transmission
configurations of control channels, for a higher control reliability,
and (b) the mandatory allocation of control channel search space for
each active device, for continuous monitoring and detection of new
resource grants, respectively. With extreme capacity demanding applications,
such as XR, the offered spectral efficiency of the network is highly
degraded by the control channels, and subsequently, the network is
only able to support a small number of XR devices. 

Current control channel design mindset ignores the fact that not all
XR traffic is equally important, and hence, not equally impacting
the end user experience. For instance, for view-port dependent XR
streaming, packets that are contributing to the streaming portion
in the pose directions are much vital than others contributing to
the streaming edges. Thus, those must be reliably and rapidly transmitted
and successfully decoded. However, with existing specifications, the
control information for such multi-priority traffic flows are transmitted
the same conservative way, leading to a larger control channel resource
overhead per device, and therefore, reducing the total number of supported
users due to resource starvation. 
\begin{figure}
\begin{centering}
\includegraphics[scale=0.43]{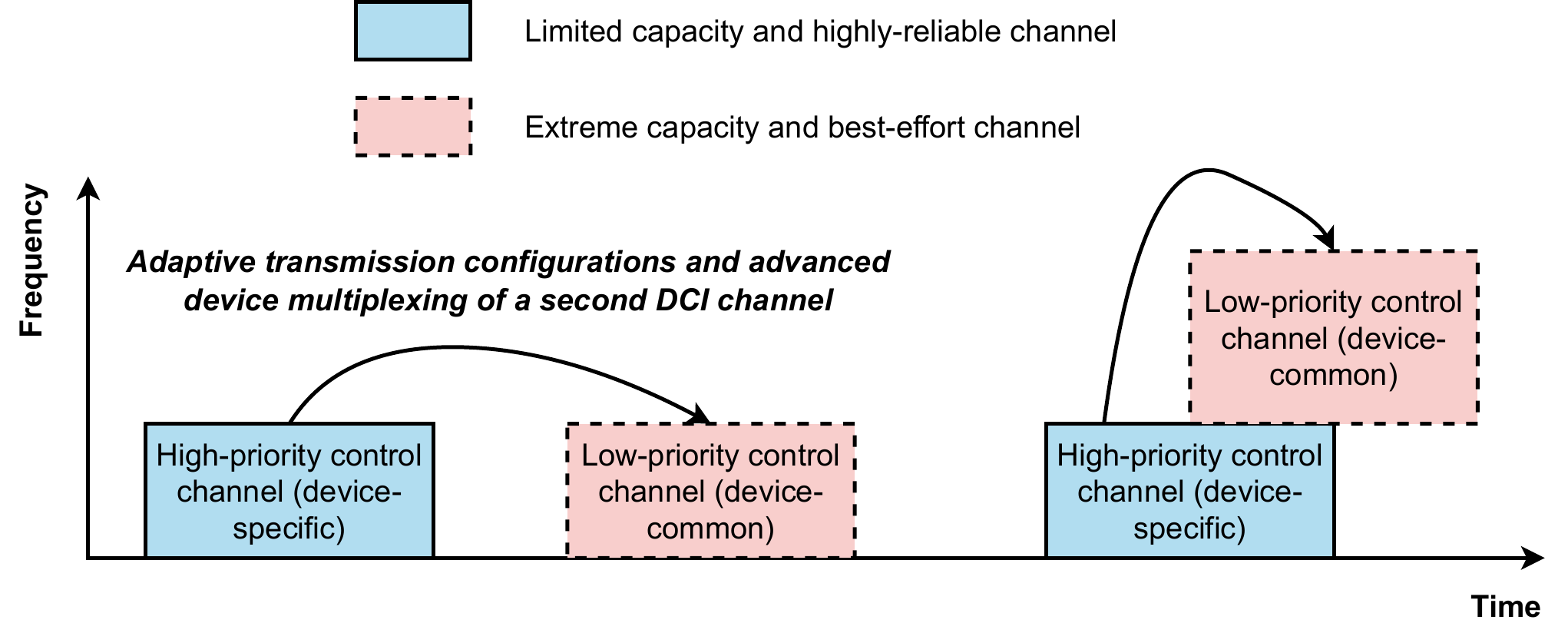}
\par\end{centering}
\centering{}\caption{Traffic-aware control channel design.}
\end{figure}
 
Thus, as depicted by Fig. 5, a traffic-aware control channel design
is proposed for next generation cellular systems. The control channel
per device is divided into two stages. Stage-1 control channel is
device-specific, and is dynamically associated with the respective
latency and reliability critical traffic flows. Thus, stage-1 control
channel information is conservatively transmitted, i.e., transmitted
with a highly conservative modulation and coding scheme (MCS) {[}19{]}.
However, stage-2 control channel is dynamically configured to carry
control information associated with best effort traffic flows. Thus,
this allows for: (1) adopting aggressive and capacity-friendly MCS
levels for transmitting the stage-2 control channel information, and
(2) multiplexing the control information of multiple devices on the
same resource sets (i.e., best effort devices and/or best effort traffic
flow's scheduling information of critical devices), and thus, boosting
the control channel capacity. The degraded decoding reliability of
stage-2 control channels (due to aggressive MCS adoption and device
multiplexing), carrying the control information of best effort traffic
flows or devices, is therefore dynamically traded-off versus the enhanced
control channel capacity, detection reliability and latency of stringent
traffic flows, respectively.

To assess the performance of the proposed multi-stage control channel
design, we adopt the same simulation setup, described in Section III.A,
and simulation parameters in Table I, where dropped UEs are randomly
categorized into low and high priority devices. Thus, Fig. 6 depicts
the ECDF of the scheduled transport block (TB) sizes in KBytes, averaged
over downlink and uplink directions for all scheduled UEs, with the
existing 5G and proposed control channel designs, respectively. The
average scheduled transport block size is a vital performance indicator
of the control channel starvation since it reflects the realistic
scheduled traffic size, based on actual decoded control channel, which
are carrying the resource scheduling information. In case of a control
channel starvation, i.e., there are not resources available for transmitting
a new scheduling control channel towards a UE, the transmission of
the respective transport blocks will be delayed until sufficient control
channel sources are freed. As shown by Fig. 6, the proposed traffic-aware
control channel design obviously enables the control channel to simultaneously
schedule more UEs, and thus, a larger average transport block sizes,
compared to existing traffic-neutral control channel. The achievable
capacity gain is attributed to the low-priority inter-UE multiplexing
of the proposed control channel framework (i.e., scheduling more UEs),
while high-priority UEs are still offered dedicated control channels.
It is also observed that the achievable scheduling gain over the lower-percentiles,
i.e., cell-edge devices, is much less than that is of the higher-percentiles,
i.e., cell-center UEs. This is due to the degraded decoding reliability
of the low-priority control channels (which is particularity impacting
the coverage-poor cell-edge UEs), and hence, cell-edge UEs may miss
detecting the scheduling information and accordingly, demanding another
control channel transmission, which leads to delaying the transmission
of the corresponding transport blocks. This behavior can  be resolved
by enforcing cell-edge UEs to be always part of the high-priority
UE group, based on the received coverage level. 

\begin{figure}
\begin{centering}
\includegraphics[scale=0.45]{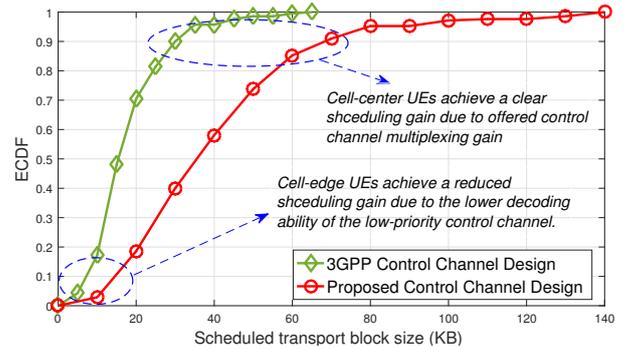}
\par\end{centering}
\centering{}\caption{ECDF of the scheduled TB sizes (KBytes).}
\end{figure}

\begin{figure*}
\begin{centering}
\includegraphics[scale=0.5]{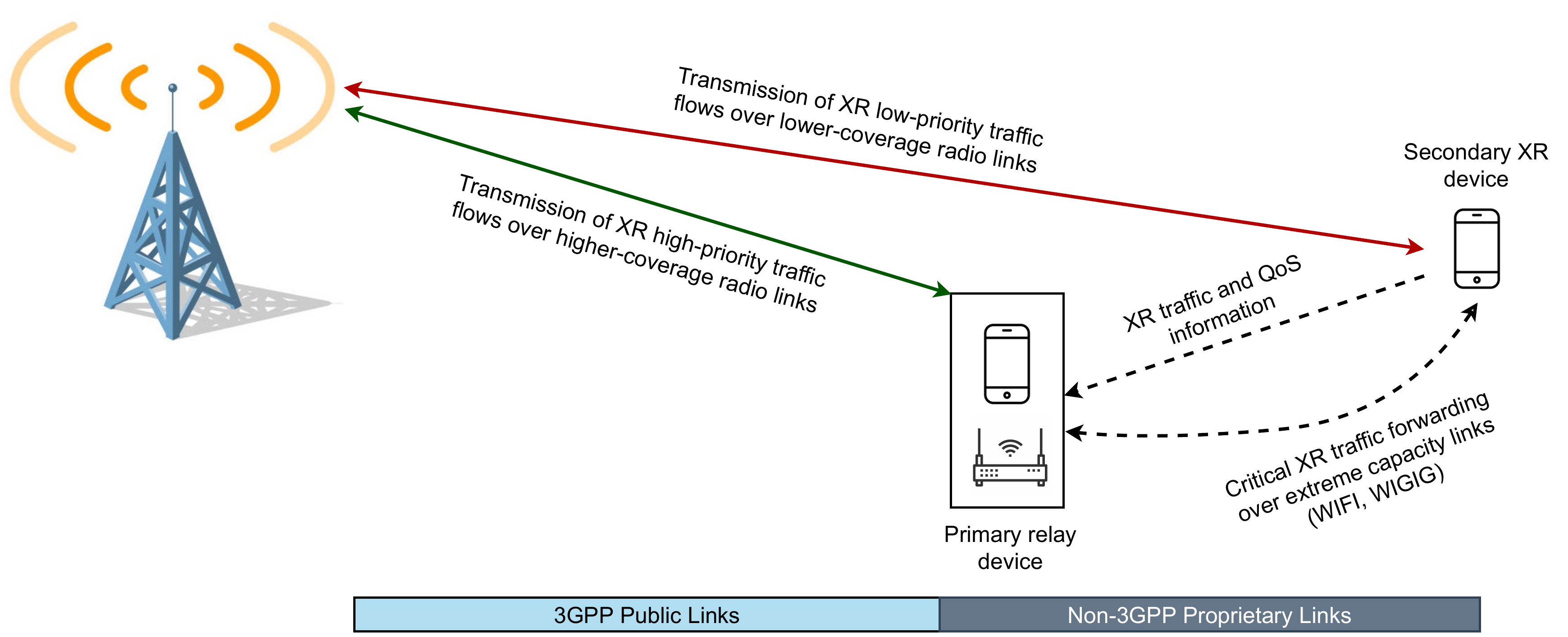}
\par\end{centering}
\centering{}\caption{Collaborative device aggregation for dynamic XR capacity boosting
and offload.}
\end{figure*}

\subsection{Collaborative Device Aggregation For Capacity Boosting}

The majority of XR services demands extreme perceived capacity, due
to the augmentation of the immersive video streaming in both downlink
and uplink directions, respectively. This calls for novel and collaborative
capacity-boosting procedures, specifically designed for XR traffic.
Device aggregation {[}20{]} is introduced to allow devices, in proximity
of each other, to coordinate and dynamically perform radio functions
over proprietary and non-3GPP links, e.g., WIFI, and WIGIG {[}21{]}.
Therefore, device aggregation offers multiple performance benefits
including: (a) enhanced capacity in case the primary device, aggregating
traffic from in-proximity secondary devices, has better radio conditions
to the network than secondary devices, and (b) power/energy saving
of the secondary devices in case the execution of a subset of the
radio functions is delegated to a nearby primary device. Device aggregation
is typically transparent to the network, i.e., the network is not
aware about the formation of the device aggregation setup. However,
such full network transparency may limit the sustainability of the
primary devices to perform efficient traffic and function aggregation.
For instance, with network-unaware device aggregation, primary devices
have to monitor and attempt decoding the independent downlink control
channels of each of the connected secondary devices, which is battery
consumption inefficient. 

As can be clearly observed, device aggregation can be vital to XR
adoption success since the extreme demanded capacity can always be
handled over higher capacity links, e.g., high capacity 3GPP radio
links to primary devices, and extreme capacity non-3GPP links from
primary to intended secondary devices. However, this requires the
efficient sustainability of the processing and battery capacities
of the primary devices, and therefore, partial network awareness can
greatly enhance the overall device aggregation performance. 

As shown by Fig. 7, an exemplary network-aware device aggregation
procedure is introduced, for XR downlink traffic delegation and offload.
The traffic-heavy XR flows can be dynamically distributed across multiple
direct and indirect radio links towards intended secondary XR devices.
Specifically, the critical downlink traffic flows are transmitted
towards the primary devices over the 3GPP radio interface, and accordingly,
being forwarded towards secondary XR devices over non-3GPP and extreme-capacity
links. However, the best effort downlink traffic flows can be transmitted
directly towards the secondary devices over the lower-capacity radio
links. Obviously, the critical traffic flows are always served on
faster radio links; although, without stressing the processing and
energy capacities of the primary devices, due to only delegating the
critical flows to primary device's links while other non-significant
flows are transmitted over the original radio links. This certainly
requires the partial network awareness about which downlink flows,
of the secondary XR devices, to transmit towards the primary devices
instead, utilizing the better channel conditions and transmission
configurations, e.g., MCS, of the primary devices.

\section{Concluding Remarks }

In this work, the state-of-the-art 3GPP specifications, on emerging
extended reality (XR) service integration into the 5G new radio, are
surveyed. The key performance challenges are identified and presented,
spanning the major 5G radio aspects such as the radio resource scheduling,
control channel design, performance reporting and exchange procedures.
For the efficient support of the XR service class into 5G and beyond
systems, several radio design improvements have been proposed, and
evaluated by extensive system level simulations, which suggest upgrading
the current 3GPP specification mindset, through introducing XR programmable
and application-dependent performance reporting, traffic-aware control
channel design, and collaborative device aggregation, respectively.
A detailed XR study will be performed in a future work, considering
other radio aspects of XR mobility, XR power saving, and XR scheduling,
respectively. 

The key recommendations and insights of this paper are as follows:
(1) The emerging XR service class does not fit within existing 5G
native quality of service classes (QoS), e.g., URLLC and eMBB, due
to the stringent and joint performance requirements on reliability,
latency, and capacity, respectively, leading the efficient support
of XR services to be further challenging, (2) XR applications, even
those belonging to the same XR service category (virtual reality,
augmented reality, and mixed reality) require a diverse set of QoS
targets, and utilize various traffic profiles. However, the real-time
and application-specific demands are not typically handled over the
radio interface, leading to highly inefficient radio operations, (3)
Thus, current 3GPP specification mindset should be upgraded to introduce
true application-awareness mechanisms over the next-generation radio
interfaces, and (4) Finally, based on presented simulation results,
solutions for programmable and application-native performance reporting,
traffic-aware control channel designs, collaborative device and traffic
aggregation become vital in supporting the stringent QoS requirements
of emerging XR services.

\end{document}